\documentclass[aps,twocolumn,preprintnumbers,amsmath,amssymb]{revtex4}

\usepackage{graphicx}
\usepackage{dcolumn}
\usepackage{bm}

\begin{document}

\preprint{}

\title{Stability of ferroelectric ice}

\author{Toshiaki Iitaka}
\email{tiitaka@riken.jp}
\homepage{http://www.iitaka.org/}
\affiliation{
Computational Astrophysics Laboratory, RIKEN Advanced Science Institute(ASI) \\
2-1 Hirosawa, Wako, Saitama 351-0198, Japan}

\date{\today}

\begin{abstract}
We theoretically study the stability conditions of the ferroelectric ice of the $Cmc2_{1}$ structure, which has been considered, for decades, one of the most promising candidates of the low temperature proton-ordered phase of pure ice Ih.  It turned out that the $Cmc2_{1}$ structure is stable only with a certain amount of dopant and the true proton-ordered phase of pure ice Ih remains to be found at lower temperature. Implication for spin ice is mentioned.
\end{abstract}

\maketitle


Water is a common molecule in the universe, found on the earth and other solar/extrasolar planets\cite{Tinetti2007,Charbonneau2009}.  The solid form of water, ice, is known to have an extremely rich phase diagram despite its simple molecular structure\cite{Petrenko1999}. The complexity comes from configurations of its hydrogen bond network. Among its many phases ice Ih is the most abundant on earth. It is characterized by hexagonal symmetry and disordered tetrahedral hydrogen bonds which satisfy the {\em ice rules}\cite{Pauling1935}.  The residual entropy $S_0$ of 3.5 (J/mol K) due to the disorder is observed when ice Ih is cooled toward the absolute zero.  At a glance this may seem to contradict the third law of thermodynamics, which states that entropy will approach zero as temperature approaches absolute zero, which is why the low temperature proton-ordered phase of ice Ih is long sought\cite{Bramwell1999}.
The existence of proton-ordered ice in space \cite{McKinnon2005,Fukazawa2006} and its role in planet formation \cite{Iedema1998,Wang2005} have also been discussed by several authors.
 It has been experimentally found that, when doped with salt such as KOH (or placed in an electric field), ice Ih transforms to "ice XI" below 72 K \cite{McKinnon2005,Fukazawa2006,Iedema1998,Wang2005,Kawada1972,Matsuo1986,Li1995,Abe2005,Arakawa2009}, which is a proton-ordered, ferroelectric crystal with space group $Cmc2_1$. The traditional view is that the pure $Cmc2_1$ structure is thermodynamically stable at low temperature and the role of a dopant is that of a {\em catalyst}.
In this Letter, we examine the stability conditions of the $Cmc2_1$ structure, which has been considered for decades one of the most promising candidates for a low temperature proton-ordered phase of pure ice Ih\cite{McKinnon2005,Fukazawa2006,Iedema1998,Wang2005,Kawada1972,Matsuo1986,Li1995,Abe2005,Arakawa2009,Pisani1996,Casassa2005,Erba2009,Kuo2003,Hirsch2004,Morokuma1984,Singer2005,Kuo2005,Itoh1998,Neto2006}.  A simple model based on first principles calculation suggests that the $Cmc2_1$ structure is stable only with a certain amount of dopant, and the true proton-ordered phase of pure ice Ih remains to be found at lower temperature.


%
%
\begin{figure}
\begin{center}
\resizebox{0.4 \textwidth}{!}{\includegraphics{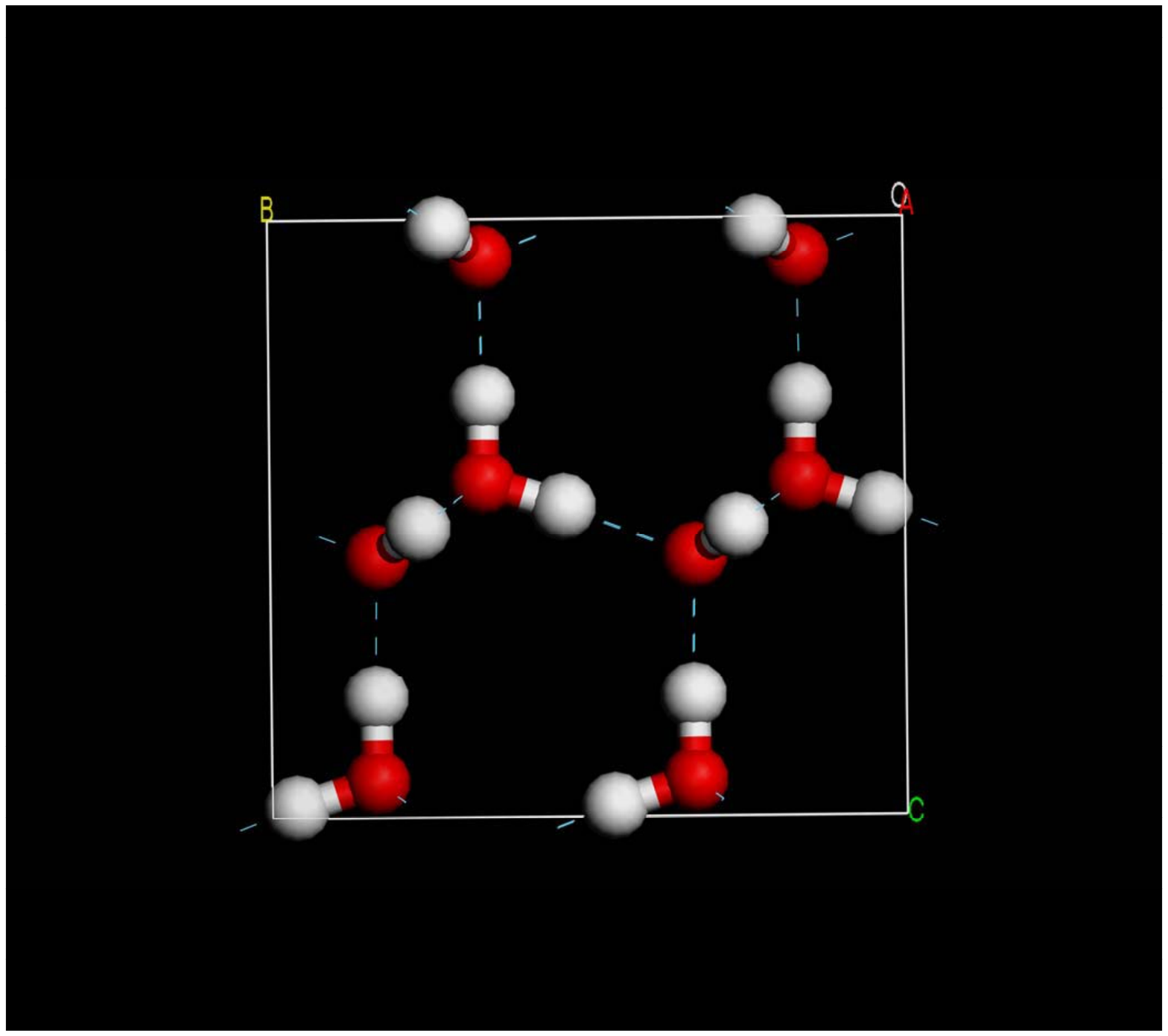}}
\end{center}
\caption{
\label{fig:unitcell}
The figure shows the $Cmc2_{1}$ hydrogen bond configuration, one of the 16 proton-ordered structures with eight water molecules in orthorhombic unit cell\cite{Hirsch2004}.
}

\end{figure}

\begin{table}
\begin{tabular}{r|r|rrr|r}
\hline 
 \# & $E_{KS}(n)-E_{KS}(1)$ & $P_{0x}$ & $P_{0y}$ & $P_{0z}$ & ($2\pi P_{0}^2/\epsilon) \Omega$ \\
\hline
1  & 0           & 0.0000   & 0.0000  & 0.0035  & 0.0720  \\
2  & 0.001196435 & 0.0000   & 0.0000  & 0.0000  & 0.0000  \\
3  & 0.001101339 & 0.0042   & 0.0000  & 0.0000  & 0.1090  \\
4  & 0.000676559 & 0.0000   & 0.0000  & -0.0034 & 0.0716  \\
5  & 0.000384392 & 0.0000   & 0.0000  & 0.0035  & 0.0720  \\
6  & 0.001494462 & 0.0000   & 0.0000  & 0.0000  & 0.0000  \\
7  & 0.000797285 & 0.0000   & 0.0000  & 0.0000  & 0.0000  \\
8  & 0.000290129 & 0.0043   & 0.0000  & -0.0034 & 0.1818  \\
9  & 0.000487221 & 0.0000   & -0.0037 & -0.0034 & 0.1539  \\
10 & 0.000662458 & 0.0021   & -0.0037 & 0.0000  & 0.1095  \\
11 & 0.000527363 & 0.0021   & 0.0000  & 0.0000  & 0.0277  \\
12 & 0.001151043 & 0.0021   & 0.0000  & 0.0000  & 0.0272  \\
13 & 0.001145400 & -0.0021  & 0.0000  & 0.0000  & 0.0273  \\
14 & 0.000383602 & 0.0000   & 0.0000  & 0.0000  & 0.0000  \\
15 & 0.000665316 & 0.0000   & -0.0037 & 0.0000  & 0.0818  \\
16 & 0.000337611 & 0.0021   & 0.0000  & -0.0034 & 0.0994  \\
\hline
\end{tabular}
\caption{Kohn-Sham total energy $E_{KS}$, polarization $\vec{P}_{0}$ and the macroscopic electrostatic energy $(2\pi P_{0}^2/\epsilon) \Omega$ (in Hartree atomic units) of the crystallographically inequivalent 16 proton-ordered structures of ice Ih for an orthorhombic unit cell with eight water molecules are listed where $\epsilon=1.8$ is the dielectric constant.}
\end{table}

The proton-ordered phase of ice Ih has also been theoretically studied and a number of works published \cite{Pisani1996,Casassa2005,Erba2009,Kuo2003,Hirsch2004,Morokuma1984,Singer2005,Kuo2005,Itoh1998,Neto2006}. Among them Kuo and Singer\cite{Kuo2003} and Hirsch and Ojamae\cite{Hirsch2004} identified crystallographically inequivalent 16 proton-ordered structures for an orthorhombic unit cell with eight water molecules (Fig.~\ref{fig:unitcell}). This classification enabled systematic comparison of hydrogen bond configurations. Following this classification we have calculated the Kohn-Sham total energy $E_{KS}$, permanent polarization $\vec{P}_{0}$, and dielectric constant $\epsilon$ of the 16 structures (Table~1) by using the first principles electronic structure calculation (see below for details). The phonon contribution to the energy is neglected. The calculated energies agree with Hirsch's results.
%
%
Structure~1 (space group $Cmc2_{1}$) is ferroelectric and the most stable, which agrees with experimental observations. Structure~2 (space group $Pna2_{1}$) is  an antiferroelectric crystal, which Davidson and Morokuma\cite{Morokuma1984} suggested early on as the most stable structure. However, these results were calculated for an infinite crystal under periodic boundary conditions, thus effects of the {\em macroscopic} electric field $\vec{\cal E}$ due to the surface charge were neglected \cite{Stengel2009} while the {\em local} interactions between water molecules were correctly taken into account.

%
%

\begin{figure}[h]
\begin{center}
\resizebox{0.4 \textwidth}{!}{\includegraphics{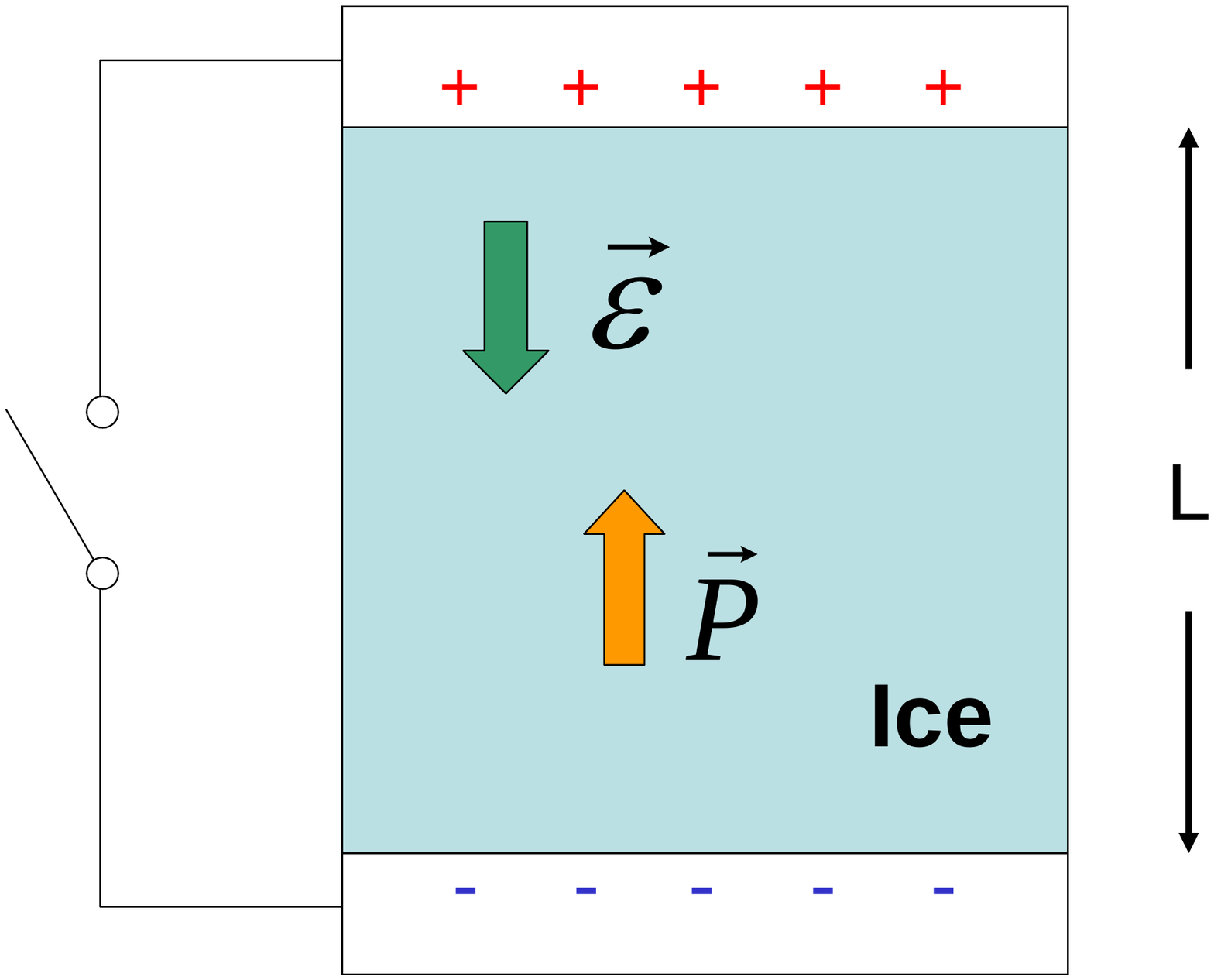}}
\end{center}
\caption{
\label{fig:model}
Model of ice crystallite: A cubic crystallite of size $L$ is sandwiched by two hypothetical electrodes that represent the surface charges due to the polarization $\vec{P}$ and due to concentration of dopant. 
}

\end{figure}

In order to evaluate the effects of the macroscopic electric field in polar crystallites, let us introduce a model with a cubic crystallite of size $L \gg \Omega^{1/3}$ sandwiched by two hypothetical 'electrodes' with charge $q=q_{pol}+q_{imp}$  and $-q$, respectively (Fig.~\ref{fig:model}) where $\Omega$ is the volume of the unit cell,  $q_{imp}$ is the doped charge, and $q_{pol}$ is the surface {\it polarization} charge due to the polarization $\vec{P}=\vec{P}_{0}+\chi \vec{{\cal E}}$ where the first term is due to the permanent dipole moment of the oriented water molecules and the second term is due to the induced dipole moment. We use Gaussian units in this Letter.
In the case of pure ice crystallite, where we have an open circuit boundary condition with $\vec{D}=\vec{\cal E}+4\pi\vec{P}=0$ or $q_{imp}=0$, the macroscopic field called {\it depolarization field}, $\vec{\cal E}=-(4\pi/\epsilon)\vec{P}_{0}$, is caused by the surface {\it polarization} charge $q_{pol}=P_{0}L^2/\epsilon$, where $\epsilon=1+4\pi\chi$ is the dielectric constant. It was evaluated by the density functional linear response theory that $\epsilon=1.8$ for our systems.

If a doped charge exists in the crystallite, it feels the force due to the macroscopic electric field $\vec{\cal E}$ and accumulates at the oppositely charged surface. Then the surface polarization charge is screened by the doped charge,
\begin{equation}
q=q_{pol}+q_{imp}=PL^2-\rho_{imp}L^3
.
\end{equation}
where $\rho_{imp}$ is the average number density of the doped charge in the crystallite. 
The minimum density $\rho_{min}$ is defined as the number density of the doped charge that screens the surface polarization charge perfectly, i.e., $\vec{\cal E}=0$ or $q=0$, 
\begin{equation}
\label{eq:rhomin}
\rho_{min}=P_{0}/L
.
\end{equation}
Dopant exceeding $\rho_{min}$ has no effect on stability in this model.
Let us define the dimensionless electric field $x$ by $\vec{\cal E}=-(4\pi/\epsilon) \vec{P}_{0} x$ or $x=1+q_{imp}/q_{0}$ where $q_{0}=P_{0}L^2$. The system with $x=1$ represents the pure system and that with $x=0$ represents the fully screened system. Since $\vec{\cal E}=0$ for the fully screened system it also corresponds to the system calculated with the conventional density functional calculation method with periodic boundary conditions, whose total energy  $E_{tot}(x=0)$ is  given by the Kohn-Sham total energy $E_{KS}$. In order to evaluate the total energy of pure system $E_{tot}(x=1)$ we calculate the work $W$ necessary for moving the doped charge $q_{imp}$ across the crystallite against the macroscopic electric field $\vec{\cal E}$. Let us define infinitesimal work $dW$ necessary for moving infinitesimal charge $dq_{imp}=q_{0}\cdot dx$ across the crystallite against the macroscopic electric field as 
\begin{equation}
dW=(dq_{imp}){\cal E}L=\frac{4\pi}{\epsilon} P_{0}^2L^3 xdx
.
\end{equation}
Then the total energy per unit cell at dimensionless surface charge $x$ becomes
\begin{eqnarray}
\label{eq:total_energy2}
E_{tot}(x)&=& E_{KS} + \frac{4 \pi}{\epsilon} P_{0}^2 \Omega \int_{0}^{x} x'dx' \nonumber \\
          &=& E_{KS} + \frac{2 \pi}{\epsilon} P_{0}^2 \Omega x^2
.
\end{eqnarray}
Here we find the main result of this paper that the $Cmc2_{1}$ structure of pure ice is unstable due to the electrostatic energy represented by the second term of eq.(\ref{eq:total_energy2}) and the ground state of pure ice should be non-polar, while the $Cmc2_{1}$ structure of doped ice is stable because the dopant acts as a stabilizer that eliminates the electrostatic energy.  All previous DFT calculations provided the total energy of doped ice $E_{tot}(x=0)=E_{KS}$, which is lower than the total energy of the pure ice $E_{tot}(x=1)$ by the electrostatic energy. The electrostatic energy for the crystallite $(2 \pi/\epsilon) P_{0}^2 L^3$ is equal to the electrostatic energy $(1/2) q_{0}^2 /C$ of a parallel plate capacitor with capacitance $C=(\epsilon/4\pi) L$.  Note that the electrostatic energy is about hundred times larger than the variation in $E_{KS}$ of various hydrogen bond configurations (see Table~1).  More precise and sophisticated treatment of finite electric field calculation in the context of the self-consistent density functional theory is found in the recent article\cite{Stengel2009}.

This electrostatic energy makes pure monodomain crystalline of typical ferroelectric materials such as $\mathrm{BaTiO_3}$ unstable.  According to the standard explanation of domain formation, the electrostatic energy of these materials is lowered by forming domains of polarization where the polarization in the half of the domains is reversed to reduce the effective polarization $P_{eff}$ at the cost of domain wall formation energy\cite{Nambu1994}. Domain size is determined by the balance of the gain in the electrostatic energy and the loss in domain formation energy. As the result the stable state of the pure ferroelectric crystalline breaks into small domains with alternating polarization instead of being one monodomain of crystal structure.

In our case of pure ice crystallite, the electrostatic energy is completely eliminated by changing the hydrogen bond configuration from polar to non-polar while leaving the oxygen lattice intact. As the result the most stable state will be one non-polar monodomain structure. This hydrogen bond reconfiguration, however, can be regarded as an extreme case of domain formation. For example, anti-ferroelectric structure $Pna2_{1}$ can be considered as ferroelectric domains with alternating directions in molecular scale. The domain size can become so small because the cost of domain formation, or the energy difference among hydrogen bond configurations, is extremely smaller than the gain in electrostatic energy (See Table~1).

%
%

\begin{figure}[h]
\begin{center}
\resizebox{0.4 \textwidth}{!}{\includegraphics{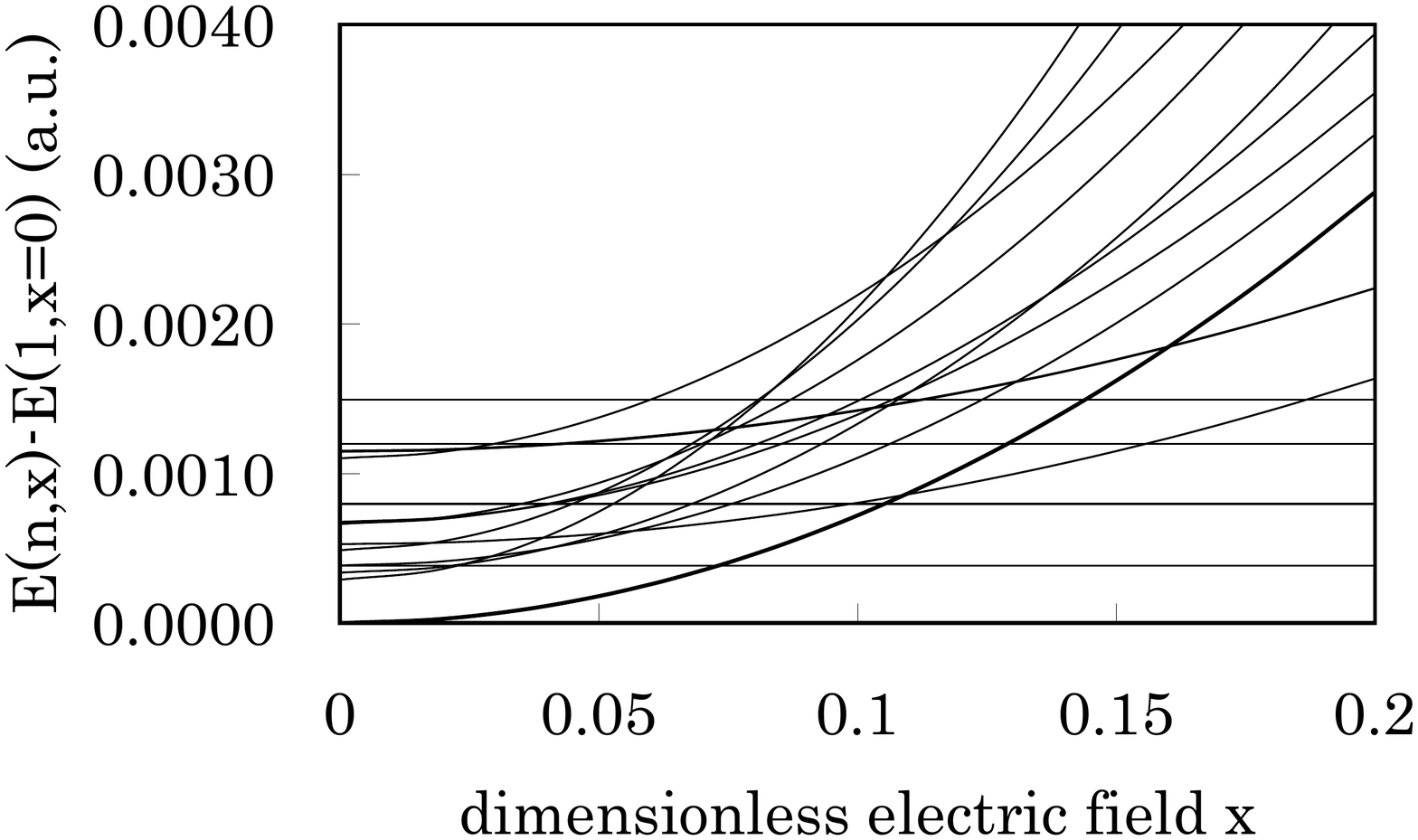}}
\end{center}
\caption{
\label{fig:energy}
Kohn-Sham total energy of the n-th Structure compared to the energy of the $Cmc2_{1}$ structure at $x=0$, $E_{tot}(n,x)-E_{tot}(1,x=0)$ (in Hartree atomic units) as a function of the dimensionless electric field $x={\cal E}/(4\pi P_{0}/\epsilon)$ where $\epsilon=1.8$ is the dielectric constant.
}

\end{figure}

Fig.~\ref{fig:energy} shows the total energy $E_{tot}(x)$ as a function of dimensionless surface charge $x$. The $Cmc2_{1}$ structure remains the most stable only up to $x=0.08$. 
The total energies of the four non-polar structures (No.~2, No.~6, No.~7, No.~14) do not change with doping within this model. Therefore the most promising candidate of the proton-ordered form of pure ice Ih is the Structure~14 (space group $P21$) which has the lowest total energy among the four non-polar structures. Non-polar configurations in the orthorhombic unit cell containing up to 64 molecules were explored by randomly generating hydrogen bonds satisfying ice rules and then optimizing the geometry with density functional calculation \cite{epaps}. Preliminary results do not show any non-polar configuration which has lower energy than the Structure~14. Random networks of hydrogen bond satisfying the ice rules appear about 200K above the energy of the $Cmc2_{1}$ structure. The phase transition temperature of doped ice, 72K, is much lower than 200K probably because of huge number of random hydrogen bond networks.  Since the total energy of the Structure~14 is 100K higher than the $Cmc2_{1}$ structure the transition temperature of pure ice is estimated to be around 36K. 

In the case of fully doped ice, the electrostatic energy in eq.(\ref{eq:total_energy2}) is completely eliminated by the dopant and the stable state of the system becomes the $Cmc2_1$ structure.  The specimen may consist of randomly oriented crystallites (or domains) with the size constrained by eq.( \ref{eq:rhomin}). The size of crystallite is not yet experimentally well determined but it is estimated from eq.(\ref{eq:rhomin}) to be as large as $L=1 \mu m$ for a typical dopant density of $\rho=0.001 (mol/l)$. Even in such a case the above argument for an isolated crystallite is considered to be valid. The effects of the electric fields originating from other domains will be negligible because the surface charge is screened and the crystallites orient randomly. 

Even the purest water does not consist only of $H_{2}O$ molecules but also contains hydronium ions ($H_{3}O^{+}$) and hydroxide  ions ($OH^{-}$) due to autoionization of the water molecules. In order to see if these ions taken into the ice can stabilize the $Cm2_1$ structure, let us calculate the minimum size of domain $L_{min}=P_{0}/\epsilon\rho$ to be stabilized by autoionization. Assuming ions of density $10^{-7}$ (mol/{\it l}) in water at standard condition are incorporated into the crystallite, the minimum domain size $L_{min}$ is estimated to be as large as 1 (cm). Nucleation of domains of such large size at once would be difficult.  Further, the dissociation constant in ice is orders of magnitude smaller than that in water \cite{Eigen1958}. 


The details of the numerical calculations are as follow. The density functional electronic structure calculations were performed with ABINIT codes \cite{abinit} based on the plane wave basis set, norm conserving pseudopotential, and GGA density functional according to Perdew, Burke and Ernzerhof (PBE)\cite{pbe}. The Brillouin zones were sampled with the Monkhorst-Pack k-points \cite{Monkhorst1976} with $6 \times 3\times 3$ mesh. The cut-off energy of plane wave basis was set to be 50 (Hartree). The positions of atoms were fully optimized so as to minimize the total energy. The initial atomic configurations were adopted from Hirsch's table \cite{Hirsch2004}. The unit cell was not optimized because the change in stress tensor due to the hydrogen bond configurations turned out to be very small. The dielectric constant was calculated by using the density functional linear response theory\cite{dflr}. The permanent polarization was calculated following the Berry phase theory\cite{King-Smith1993}.


In summary we theoretically studied the stability conditions for the ferroelectric ice of the $Cmc2_{1}$ structure. It turned out that the $Cmc2_{1}$ structure is stable only with a certain amount of dopant. The true proton-ordered phase of pure ice Ih should be non-polar and remains to be found at lower temperature.  We proposed the Structure~14 (space group $P21$) as a promising candidate. In the formation of ferroelectric ice ($\vec{P}_{0} \ne 0$), dopant acts not only as a catalyst but also as a stabilizer eliminating the second term of eq.(\ref{eq:total_energy2}) by screening the surface polarization charge. Contrarily, a dopant acts only as a catalyst in the formation of non-polar proton-ordered ices ($\vec{P}_{0}=0$) \cite{Salzmann2006} because the macroscopic electrostatic energy is zero even without dopant. 
 The recent experimental discovery of antiferroelectric ice XV \cite{Salzmann2009} in spite of the theoretical prediction of ferroelectric structure \cite{Knight2005,Kuo2006} might be relevant to our model. Since the surface polarization charge is screened by doped charges or the hydrogen bond network is reconstructed to non-polar structures, the astronomical implications of the strong electric field produced by ferroelectric ice \cite{Iedema1998,Wang2005} sound unlikely.

 It has long since been clear that there is a similarity between water ice and spin ice\cite{Anderson1956}. In spin ice, magnetic monopoles\cite{Castelnovo2008,Fennell2009} interacting with an external magnetic field\cite{Morris2009,Kadowaki2009} have been observed in neutron scattering experiments. In close analogy to the ferroelectric water ice, the effect of surface magnetic charge may also be important for crystallites of ferromagnetic spin ice.

\section*{Acknowledgment}
This work was supported by KAKENHI (No.20103001-20103005 and No.19310083) from MEXT of Japan.  Numerical calculations were conducted on the RIKEN Cluster of Clusters (RICC). Many thanks to David W. Chapmon for editing and stylistic revision of this Letter.



%



\end{document}


\preprint{}

\title{"Stability of ferroelectric ice"\\
Supplemental Materials}

\author{Toshiaki Iitaka}
\email{tiitaka@riken.jp}
\homepage{http://www.iitaka.org/}
\affiliation{
Computational Astrophysics Laboratory, RIKEN Advanced Science Institute(ASI) \\
2-1 Hirosawa, Wako, Saitama 351-0198, Japan}

\date{\today}


\maketitle

\renewcommand{\tablename}{S}
\appendix

\section*{Crystal Structure Prediction for Hydrogen Bond Materials}
We searched the hydrogen bond configurations in the 1x1x1, 2x1x1, 2x2x1 and 2x2x2 supercells of the 8 molecule orthorhombic unit cell \cite{Hirsch2004} by using crystal structure prediction program for hydrogen bond materials. We adopted multi-energy scale strategy by using empirical potential to select low energy configurations and first principles calculation to distinguish small energy differences among them. The initial configuration was generated by choosing the orientation of each water molecule randomly from possible six orientations, then the total energy was minimized in terms of water orientation by using Metropolis method for simulated annealing and PSC/E model potential \cite{Berendsen1987} for water-water interaction. The resulting configuration was accepted if the polarization is zero and the total energy is less than cut-off energy to select hydrogen bond configurations that satisfy the ice-rules. We collected 100 such low energy configurations for each supercell. The crystallographic equivalence of the accepted configurations was not checked. Then the atomic positions of the accepted configurations were further optimized by using density functional total energy calculation program VASP\cite{vasp} and the distribution of the total energy was derived. (Fig.~S1). The energy distributions were not sampled perfectly uniform but we believe they provide useful information on the energy distribution of hydrogen bond configurations.

\begin{figure}[h]
\begin{center}
\resizebox{0.5 \textwidth}{!}{\includegraphics{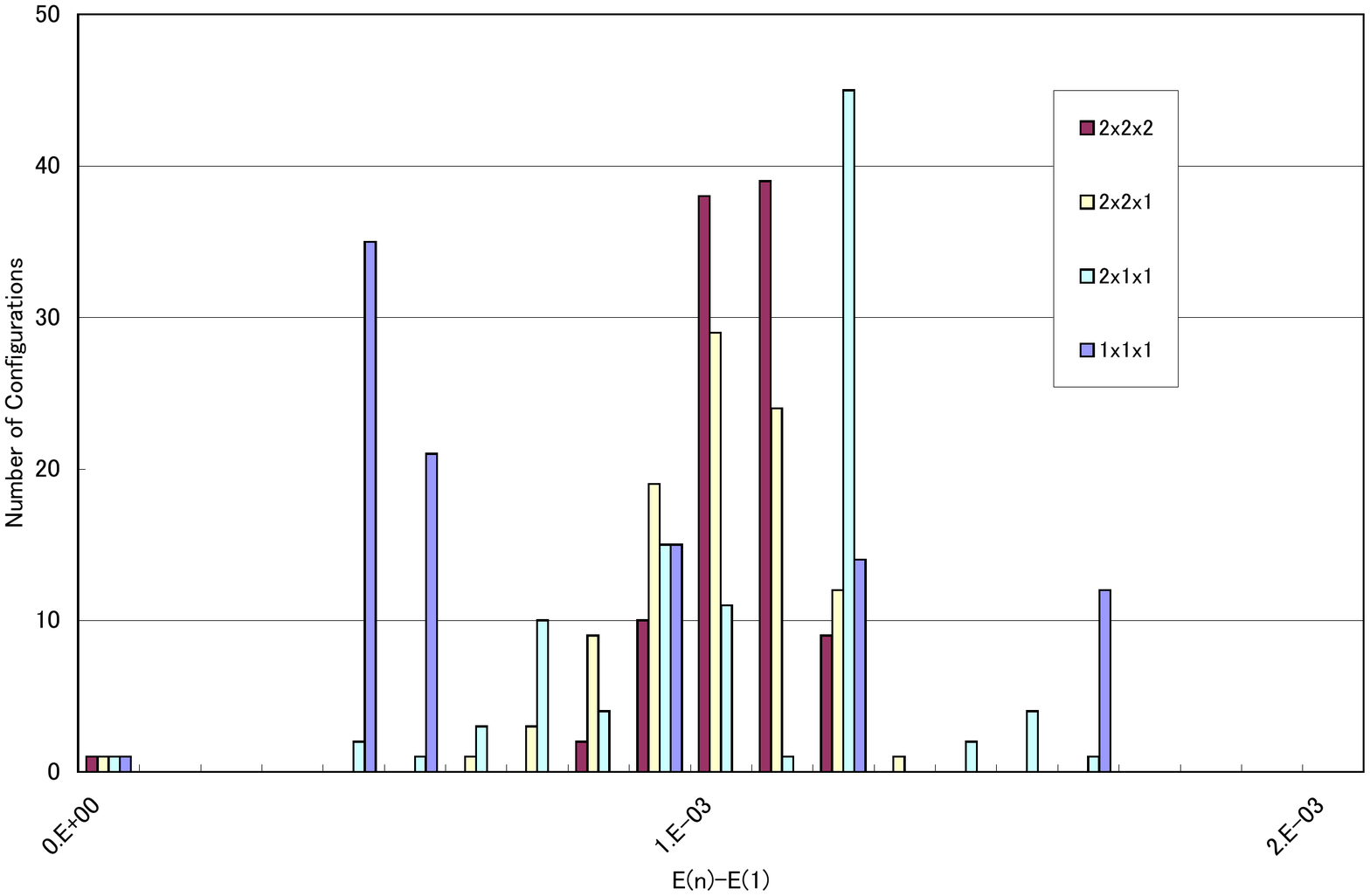}}
\end{center}
\caption{
\label{fig:energy}
Histogram of non-polar hydrogen bond configurations with respect to the total energy of the n-th Structure compared to the energy of the $Cmc2_{1}$ structure at $x=0$, $E_{tot}(n,x=0)-E_{tot}(1,x=0)$ (in Hartree atomic units).
}
\end{figure}


\clearpage

\section*{Structure No. 14}

The Structure No.~14 has the cell dimensions of $a=7.7808, b=7.33581, c=4.49225$ in the unit of Angstrom and belongs to  crystallographic space group $P21$. The fractional coordinates are listed in Table S I.

\begin{table}[h]
\begin{tabular}{lccc}
\hline 
 atom & $x$ & $y$ & $z$ \\
\hline
H1     &     0.91530  & 0.19583  & 0.24980 \\
H2     &     0.79656  & 0.02001  & 0.25135 \\
H3     &     0.52649  & 0.51741  & 0.06879 \\
H4     &     0.70263  & 0.51818  & 0.24625 \\
H5     &     0.02160  & 0.98234  & 0.92673 \\
H6     &     0.02213  & 0.98236  & 0.57414 \\
H7     &     0.41772  & 0.30443  & 0.74775 \\
H8     &     0.47740  & 0.47955  & 0.56842 \\
O9     &     0.91800  & 0.06071  & 0.25039 \\
O10    &     0.58290  & 0.56414  & 0.25143 \\
O11    &     0.08451  & 0.93553  & 0.75071 \\
O12    &     0.41707  & 0.43952  & 0.75151 \\
\hline
\end{tabular}
\caption{
Fractional coordinates of Structure No.14. 
}
\end{table}
%
